\documentclass[aps,pra,floatfix,showpacs,twocolumn]{revtex4}
\usepackage{amsmath}
\usepackage{graphicx}
\def\prn#1{{\left(#1\right)}}

\newcommand{\abs}[1]{\left|#1\right|}

\def\bra#1{{\left\langle#1\right\vert}}
\def\ket#1{{\left\vert#1\right\rangle}}

\def\prn#1{{\left(#1\right)}}
\def\brk#1{{\left[#1\right]}}

\def\abs#1{{\left|#1\right|}}

\newcommand{\oberlin}{Department of Physics and Astronomy, Oberlin College, Oberlin, OH 44074, USA}

\begin{document}
\title{Measurement of the $4\: S_{1/2} \rightarrow 6 \: S_{1/2}$ transition frequency in atomic potassium via direct frequency comb spectroscopy}
\author{J.E.\ Stalnaker}\email{jason.stalnaker@oberlin.edu}
\author{H.M.G.\ Ayer}
\author{J.H.\ Baron\footnote{Present address: Department of Physics, Harvard University, Cambridge, MA 02138}}
\author{A. Nu\~nez}
\author{M.E.\ Rowan$^\dagger$}
\affiliation{\oberlin}

\date{\today}
\begin{abstract}
We present an experimental determination of the $4 \: S_{1/2} \rightarrow 6\: S_{1/2}$ transition frequency in atomic potassium, $^{39}$K, using direct frequency comb spectroscopy. The output of a stabilized optical frequency comb was used to excite a thermal atomic vapor.  The repetition rate of the frequency comb was scanned and the transitions were excited using step-wise two-photon excitation.  The center of gravity frequency for the transition was found to be $\nu_\textrm{cog} = 822\, 951\, 698.09(13)$ MHz and the measured hyperfine $A$ coefficient of the $6\: S_{1/2}$ state was $21.93(11)$ MHz.  The measurements are in agreement with previous values and represent an improvement by a factor of $700$ in the uncertainty of the center of gravity measurement.
\end{abstract}
\pacs{42.62.Fi, 32.70.-n}


\maketitle
\section{Introduction}
Optical frequency combs provide an intriguing light source for performing spectroscopy (for a review of direct frequency comb spectroscopy see, e.g.\ Ref.\ \cite{stowe08}).  By utilizing the direct output of a frequency comb one has access to multiple, well-defined frequencies that can be directly related to the SI definition of the second.  In many cases, this can lead to simpler experimental setups and allow for access to wavelengths that are difficult to produce via other means.

\begin{figure}[h]
\centering
\includegraphics[width=3.375in]{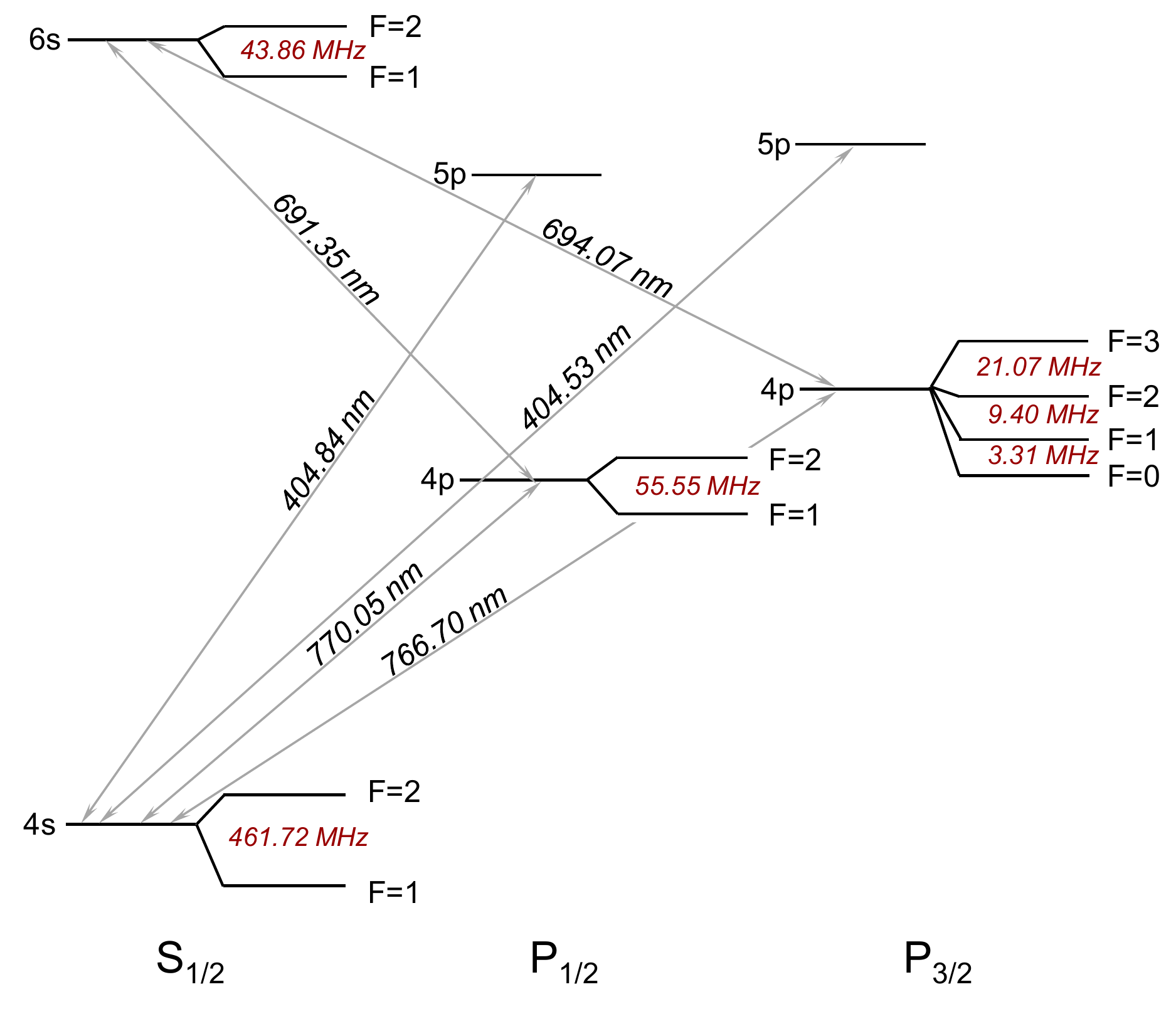}
\caption{Energy level diagram for $^{39}$K (nuclear spin $I = \frac{3}{2}$) showing the relevant energy levels and hyperfine structure.  The excitation wavelengths are indicated for the transitions used. The hyperfine structure of the states involved in the excitation are shown.  The 6 $S_{1/2}$ state is excited via stepwise excitation through the 4 $P_{J}$ intermediate states.  The excitation is detected via the fluorescence from the 5 $P_J \rightarrow 4\, S_{1/2}$ decay branch at 405~nm following the 6 $S_{1/2} \rightarrow 5\, P_{J}$ decay transition.  }\label{fig KEnergyDiagram}
\end{figure}

Here we apply the technique of velocity-selective direct frequency comb spectroscopy to excite two-photon transitions in a potassium (K) atomic vapor.  This technique has previously been applied to two-photon transitions in atomic cesium \cite{stalnaker10} and has been investigated in the context of two-photon transitions in rubidium \cite{stalnaker12}.  In this application we determine the frequencies of the 4 $S_{1/2}$ $\rightarrow$ 6 $S_{1/2}$ transitions in potassium as excited through the 4 $P_{J}$ intermediate states, see Fig.\ \ref{fig KEnergyDiagram}.  We also discuss differences that arise in the spectra due to the hyperfine structure of the intermediate states of the transitions and provide a theoretical discussion of the systematic effects associated with this spectroscopic technique.

\section{Theoretical Considerations}

In exciting two-photon transitions with the direct output of the optical frequency comb it is possible to achieve narrow, Doppler-free transitions through step-wise excitation while varying only the repetition rate of the optical frequency comb.  For the optical frequency comb used in this experiment, the output consists of $\approx 10^5$ optical modes with  frequencies
\begin{align}
  \frac{\omega_n}{2\pi} = \prn{ n\, f_\textrm{r} +f_0 },
\end{align}
where $n$ is the integer mode number of the optical mode, $f_r$ is the repetition rate frequency of the comb, and $f_0$ is the carrier-envelope offset frequency.  Stepwise excitation requires resonance for both the ground-to-intermediate state transition and the intermediate-to-final state transition.  If the comb light is split and counter-propagated through a thermal atomic source then the resonance condition is
\begin{align}
  \frac{\omega_{gi}}{2\pi} = & \prn{n_1\, f_\textrm{r} +f_0}\prn{1 +\frac{v}{c}} \\
  \frac{\omega_{if}}{2\pi} = & \prn{n_2\, f_\textrm{r} +f_0}\prn{1 -\frac{v}{c}} ,
\end{align}
where $\omega_{gi}$ is the resonant angular frequency of the ground-to-intermediate state transition, $\omega_{if}$ is the resonant angular frequency of the intermediate-to-final state transition, $v$ is the velocity class of the atoms that are excited, and $n_{1(2)}$ is the mode number of the optical mode resonant with the first (second) stage of the transition.  Here the signs of the Doppler shift terms are opposite for the two transitions since the beams are assumed to be counter propagating. For a given pair of comb modes, $n_1$ and $n_2$, there exist a repetition rate frequency and velocity class for which the resonant condition is satisfied.  If the velocity class is populated in the atomic source, atoms in the velocity class will be excited.

As the repetition rate is scanned, different pairs of comb modes come into resonance for different velocity classes, leading to a repeating excitation spectrum with excited state populations that are dependent on the velocity class that is resonant for a given pair of comb modes.  As discussed in Refs.\ \cite{stalnaker10, stalnaker12}, when the repetition rate of the optical frequency comb is larger than the Doppler distribution of the atomic sample, the resulting spectra can be well understood and the resonant frequencies can be extracted.

While it is possible to extract the resonant frequencies for both stages of the transition, the sensitivity of the experiment to the final state energy levels is significantly higher than it is to the intermediate state energy levels.  Using the resonance conditions above, we can rewrite the resonance condition in terms of the final state energy levels, $\omega_{gf}$, and the intermediate state energy levels, $\omega_{gi}$, as
\begin{align}
   \frac{\omega_{gi}}{2\pi} = & \prn{n_1\, f_\textrm{r} +f_0}\prn{1 +\frac{v}{c}} \\
  \frac{\omega_{gf}}{2\pi} = & \prn{n_1\, f_\textrm{r} +f_0}\prn{1 +\frac{v}{c}} + \prn{n_2\, f_\textrm{r} +f_0}\prn{1 -\frac{v}{c}}.
\end{align}
To understand how the energy of the intermediate state affects the repetition rate frequency where resonance occurs we consider the case where $f_0 = 0$ for simplicity.  Solving for the resonant repetition rate in this case gives
\begin{align}
  f_r = \frac{1}{2\, n_2}\prn{\frac{\omega_{gf}}{2\pi}} - \prn{\frac{n_1-n_2}{2\, n_1\, n_2}}\prn{\frac{\omega_{gi}}{2\pi}} \label{eq intStateSuppression}.
\end{align}
We note that this equation reduces to the expected result when $n_1=n_2$ and the intermediate state is exactly halfway between the ground and final states.  In addition, this relation shows that the shift in the repetition rate due to a change in the energy of the intermediate state is suppressed by a factor of $\prn{\frac{n_1-n_2}{n_1}}$ when compared to the shift resulting from the a change in the final state.   The consequence of this result for this experiment is that the hyperfine structure of the ground and final states are clearly resolved in the resulting spectra while the hyperfine structure of the intermediate state is not.  This lack of resolution of the intermediate state splittings is more significant for intermediate states whose energy levels that lie  near the midpoint of the ground-to-final state energy difference.

\section{Experimental Setup}

The frequency comb used in this experiment is described in detail in Ref.\ \cite{stalnaker12}.  Briefly, the frequency comb is generated via a mode-locked Ti:sapphire resonator, based on those discussed in Ref.\ \cite{bartels05}.  The Ti:sapphire resonator produces a coherent pulse train that spans $\approx 30$ nm centered around $\approx 780$ nm.  This light is spectrally broadened by $\approx$ 30 cm of nonlinear photonic crystal fiber with a zero GVD wavelength at 790 nm.  The output of the fiber has a power of  $\approx 200$ mW and spans $\approx 500 - 1100$ nm.  The light is filtered according to the wavelength, with the light between $650 - 1000$ nm being available for the spectroscopy and the wings of the spectrum (light with wavelengths below 650 nm and above 1000 nm) being used for the stabilization of the repetition rate frequency, $f_r$, and the carrier-envelop-offset frequency, $f_0$ through the standard $f-2f$ self referencing technique \cite{jones00}. The repetition rate frequency and the carrier-envelop-offset frequency were stabilized to synthesizers referenced to a GPS-disciplined rubidium atomic clock, resulting in a fractional instability of $\approx 10^{-11}$ in 1~s.

\begin{figure}[h]
\centering
\includegraphics[width=3.25in]{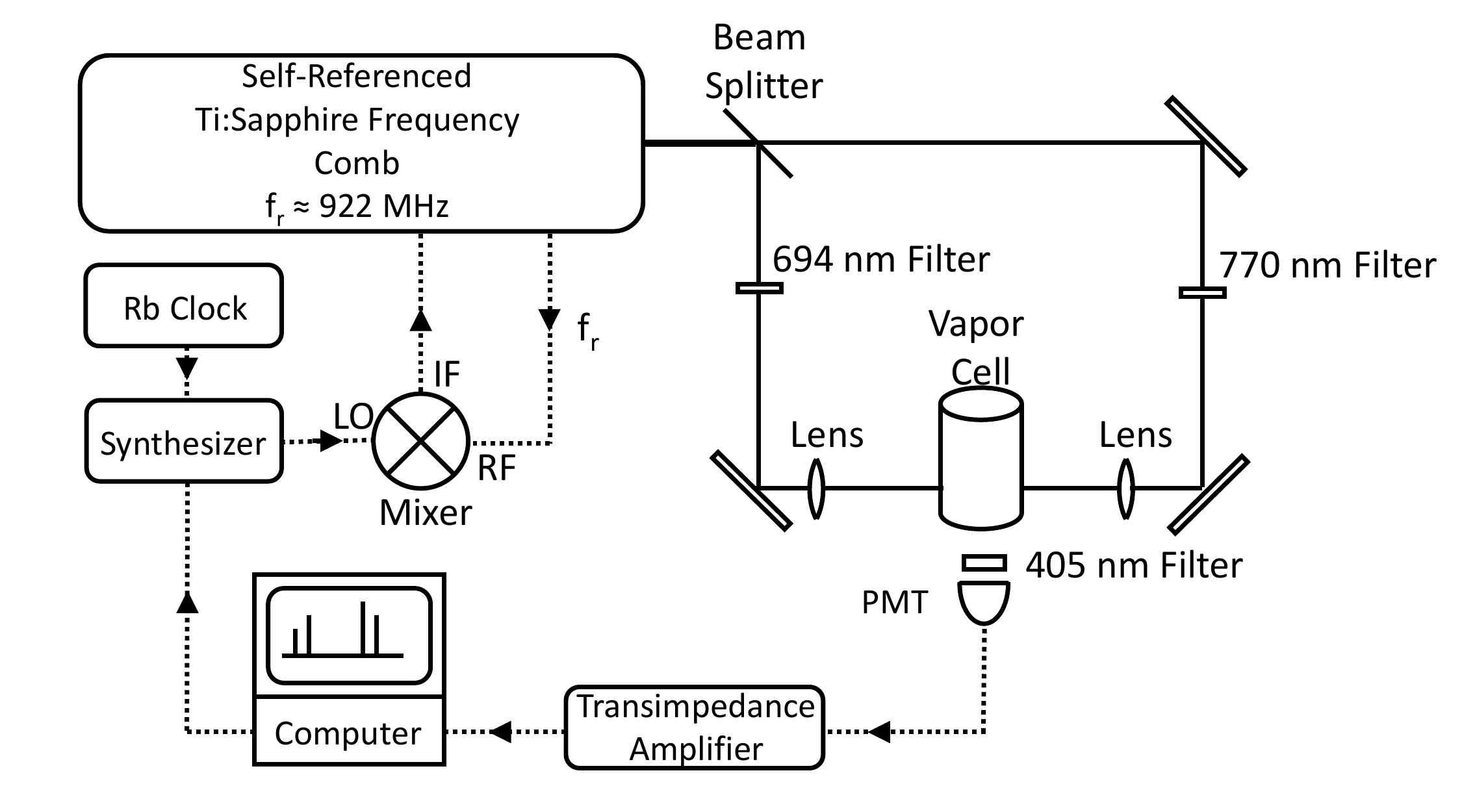}
\caption{Experimental setup.  The repetition rate frequency of the frequency comb is monitored and stabilized to a synthesizer that is referenced to a GPS-steered rubidium clock.  The frequency of the synthesizer is controlled by a computer.  The output of the frequency comb that is not used for the stabilization is split, filtered, counter-propagated, and focused into a vapor cell of potassium.  The fluorescence at 405 nm is monitored with a photomultiplier tube (PMT).  The current from the PMT is converted to a voltage by use of a transimpedance amplifier and then recorded by the computer. }\label{fig KExptSetup}
\end{figure}

Figure \ref{fig KExptSetup} shows the experimental set up.  The comb light available for the spectroscopy was split and filtered using interference filters  and counter-propagated through a potassium vapor cell with natural isotopic abundance at a temperature of $T \approx 65 ^\circ$C.  Because $^{39}$K is 93$\%$ of the natural isotopic abundance, we were only able to clearly see transitions from that isotope.    The two beams were focused using lenses with focal lengths of 10 cm.  The estimated spot size of the two beams at the focus is $\approx 50\: \mu$m.  The interference filters were selected to allow the two beams to pass light near the resonant frequency for the two transitions.  One of the filters was at $770$ nm with a $20$ nm pass band, allowing for excitation of both the $4\: P_{1/2}$ and $4\: P_{3/2}$ intermediate states (the $D_1$ and $D_2$ transitions) and the other filter was centered at $694$ nm with a pass band of 10 nm, allowing for excitation of the $4\: P_{J} \rightarrow 6\: S_{1/2}$ transitions (see Fig.\ \ref{fig KEnergyDiagram}).  The optical powers transmitted through the filters were $\approx 400\: \mu$W for the 770 nm filter and $\approx 200\: \mu$W through the 694 nm filter. The polarization of the light used for the excitation was not controlled, but is approximately linear.

The excitation to the $6\: S_{1/2}$ state was detected via fluorescence from the $5\: P_J \rightarrow 4\: S_{1/2}$ decay branch using a photomultiplier tube that had an interference filter at 405 nm with a 10 nm pass band in front of the photocathode.  The signal from the photomultiplier tube was passed through a transimpedance amplifier with a 20 kHz bandwidth.  The fluorescence spectra were acquired by scanning the frequency of the computer-controlled function generator to which the repetition rate frequency was stabilized.  The frequency was scanned in steps of 1 Hz over a range of a few hundred Hz. The range the repetition rate frequency was scanned was varied to observe the excitation from different hyperfine ground state components and different optical modes of the comb.  At each step of the repetition rate frequency the signal from the photomultiplier tube was digitized using a 16 bit analog-to-digital conversion board 5,000 times at a rate of 20,000 samples/s.  These digitizations were averaged and displayed as a function of the repetition rate frequency.  The standard deviation of the 5,000 digitizations were also recorded to provide some estimate of the statistical uncertainty associated with each average.

\section{Analysis and Results}
\subsection{Experimental and Modeled Spectra}

\begin{figure}[h]
\centering
\includegraphics[width=3.25in]{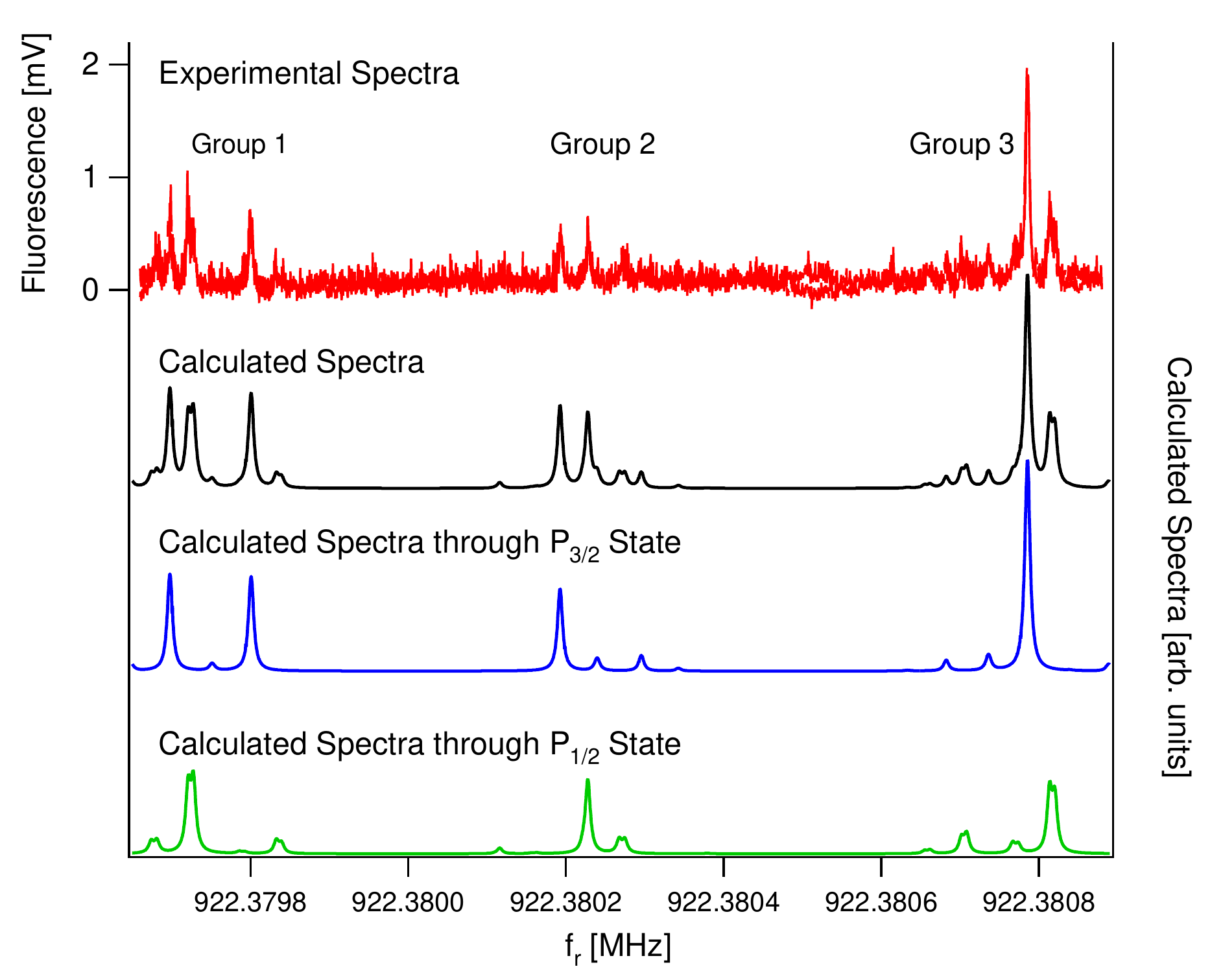}
\caption{Color Online: Experimental and modeled spectra for the $4\, S_{1/2} \rightarrow 6\, S_{1/2}$ transition.  Groups 1 and 3 correspond to excitation from the $4\, S_{1/2}\: \prn{F = 2}$ ground state with different total mode numbers.  Group 2 corresponds to excitation from the $4\, S_{1/2}\: \prn{F = 1}$ ground state.  The top trace is the experimental spectra for six scans that are concatenated together. Each scan consists of two overlapping spectra corresponding to increasing and decreasing $f_r$. The calculated spectra is shown below the experimental spectra.  The contributions to the signal from the different excitation pathways are shown in the bottom two traces. }\label{fig spectra}
\end{figure}

Figure \ref{fig spectra} shows the experimental spectra and the modeled spectra for the full range of the repetition rate used in this experiment.  The experimental data represent six different scans that were concatenated together.  Additional scans were taken over the regions where there were prominent fluorescence signals.

The modeled spectra were generated by numerically integrating the two-photon transition rate over the velocity distribution in a similar way to what is described in Refs.\ \cite{stalnaker10,stalnaker12}.  The two-photon transition rate is given by \cite{loudin83}
\begin{widetext}
\begin{small}
\begin{align}
W\prn{4\;S_{1/2} F ,\, 6\:S_{1/2} F^{\prime\prime}} = & \prn{\frac{I_1\, I_2}{4\,\epsilon_0^2\, c^2\, \hbar^4}}\prn{\frac{1}{2F+1}} \frac{\gamma_{4 P_{J^\prime}}+\gamma_{6S_{1/2}}}
{\brk{\omega_{4\:S_{1/2} F:6\: S_{1/2}F^{\prime\prime}}-\prn{\omega_1+\mathbf{k}_1\cdot\mathbf{v}}-\prn{\omega_2+\mathbf{k}_2\cdot \mathbf{v}}}^2+\prn{\frac{\gamma_{4 P_{J^\prime}}+\gamma_{6S_{1/2}}}{2}}^2} \nonumber \\
&\;\;\;\;\;\;\;\;\;\;\;\; \times \sum_{M_F,\, M_F^{\prime\prime}}
\abs{\sum_{J^\prime ,F^\prime ,M_F^\prime}\frac{\bra{6\: S_{1/2}F^{\prime\prime}
M_F^{\prime\prime}}\hat{e}_2\cdot\mathbf{d}\ket{4\: P_{J^\prime}F^\prime
M_F^\prime}\bra{4\: P_{J^\prime}F^\prime
M_F^\prime}\hat{e}_1\cdot\mathbf{d}\ket{4\: S_{1/2}F
M_F}}{\omega_{4\: S_{1/2} F: 4 \: P_{J^{\prime}}
F^{\prime}}-\prn{\omega_1+\mathbf{k}_1\cdot\mathbf{v}}-i\,\frac{\gamma_{4\: P_{J^\prime}}}{2}}}^2
, \label{eq twoPhotonProb}
\end{align}
\end{small}
\end{widetext}
where $\omega_1$ and $\omega_2$ are the angular frequencies of the excitation lasers, $M_F$, $M_F^\prime$,
and $M_F^{\prime\prime}$ are the projections of the total angular
momenta $F$, $F^\prime$, and $F^{\prime\prime}$ along the axis of
quantization, $\gamma_{n \, L_J}$
is the homogeneous line width of the state $\ket{n \, L_J}$, $\hat{e}_{1(2)}$ is the polarization vector of the first (second) light beam, and
$\omega_{n\, L_{J} F: n^{\prime}\, L^{\prime}_{J^{\prime}}F^\prime}$
is the resonant angular frequency of the transition $\ket{n\, L_{J}
\, F} \rightarrow \ket{ n^{\prime}\, L^{\prime}_{J^{\prime}}\,
F^\prime}$.  The sum over $J^\prime$ runs from $1/2$ to $3/2$, and the $F^\prime$ runs over the hyperfine levels of the intermediate state.  Linear polarization was assumed and the Wigner-Eckart theorem was used to relate the matrix elements to a reduced matrix element that is independent of the magnetic sublevels.  The reduced matrix elements in the $F$ basis were related to the reduced matrix elements in the $J$ basis using standard angular momentum relations (see, e.g., Ref.\ \cite{sobelman96}).

The two-photon transition probability was calculated for the $\approx 6$ comb modes that gave a laser frequency closest to the resonant transition frequency of each stage of the two-photon transition.  This two-photon transition probability was integrated over the Doppler distribution assuming a temperature of $T = 65^\circ$C.   The calculation was performed using the transition frequencies for the $4\: S_{1/2} \rightarrow 4\: P_{J}$ transitions as measured by Falke, \textit{et al}.\ \cite{falke06} and the measured ground state hyperfine splitting as reported by Arimondo, \textit{et al}.\ \cite{arimondo77}.  The center of gravity energy levels and the hyperfine structure splitting of the final state were varied about the previously measured values of Thompson, \textit{et al}.\ \cite{thompson83} and compared to the experimental data in order to extract the measurement of the transition energies as described in Sec.\ \ref{sec analysis}.

The spectra consists of three groups of peaks as indicated in Fig.\ \ref{fig spectra}.  These groupings correspond to transitions from different hyperfine ground states splittings and different total resonant mode numbers, $n_T = n_1 +n_2$. In particular, the group of peaks in the middle of the frequency range, group 2, correspond to the $4\: S_{1/2}\prn{F = 1}  \rightarrow 6\: S_{1/2} \prn{F^\prime}$ transitions, while the groups at the low and high frequency range of the scan correspond to the $4\: S_{1/2}\prn{F = 2}  \rightarrow 6\: S_{1/2} \prn{F^\prime}$ transitions excited with different total mode numbers ($n_{T} = 892,\, 204$ for the low frequency group and $n_T = 892,\,203$ for the high frequency group).  While the transitions are excited through both the $4\: P_{1/2}$ and $4\: P_{3/2}$ states, the suppression factor of the intermediate state frequency difference on the resonant repetition rate frequency, Eq.\ \eqref{eq intStateSuppression}, results in shifts from the intermediate fine structure splitting being smaller than the shifts arising from the ground state hyperfine splitting.  This suppression is also evident in the hyperfine splitting of the intermediate state.  While the hyperfine splitting of the $4\: P_J$ states are comparable to that of the $6\: S_{1/2}$ state, the hyperfine structure of the  $4\: P_{J}$  intermediate states are unresolved, while that of the  $6\: S_{1/2}$ hyperfine structure is fully resolved.

\begin{figure}[h]
\centering
\includegraphics[width=3.25in]{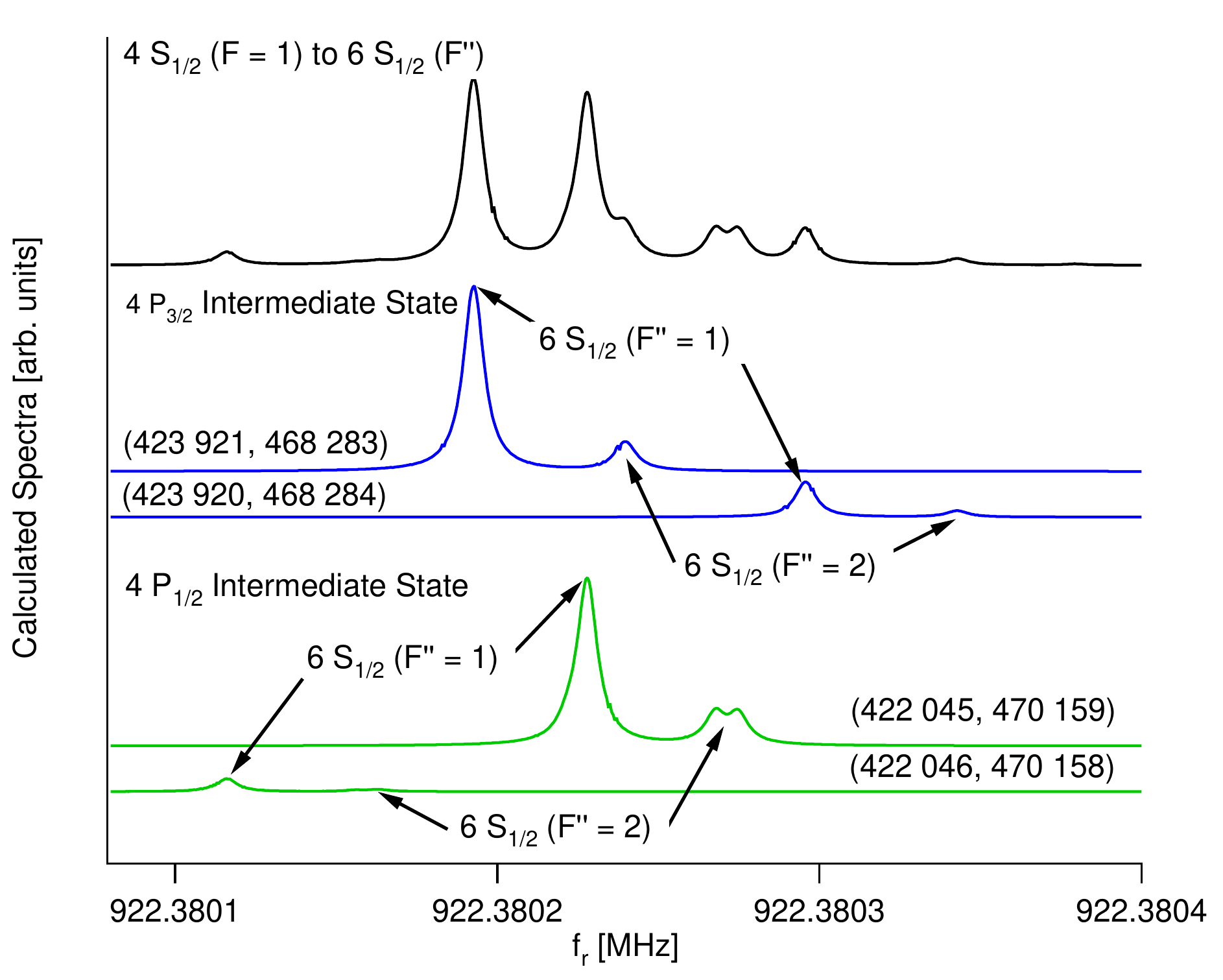}
\caption{Color Online: Modeled spectra for the $4\, S_{1/2} \prn{F=1} \rightarrow 6\, S_{1/2} \prn{F^{\prime\prime}}$ transitions.  The top trace shows the calculation including all relevant mode numbers and intermediate states.  The middle two traces (blue) show the excitation through the $4\:P_{3/2}$ intermediate state.  The two traces correspond to excitation with different pairs of mode numbers.  The bottom two traces show the excitation through the $4\:P_{1/2}$ intermediate state with different pairs of mode numbers contributing to the different peaks.  Note that in all cases the total mode number, $n_T = 892 \,204$, is the same for all of the peaks.  For a given transition, the different amplitudes are a result of different resonant velocity classes.}\label{fig group2}
\end{figure}

The effect of the velocity-selective excitation on the peak amplitudes is clearly present in the spectra.  Figure \ref{fig group2} shows an expanded view of the calculated spectra for the transitions from the $4\:S_{1/2}\prn{F=1}$ ground state, group 2 in Fig. \ref{fig spectra}.  The contributions from the different intermediate states and the different pairs of resonant optical modes of the comb are shown.  Examining the transitions through the $4\:P_{3/2}$ intermediate state, the two middle traces in Fig.\ \ref{fig group2}, we see that the same transition is excited by different pairs of optical modes at different repetition rate frequencies and that the amplitudes of the peaks are different.  This is a result of the different resonant velocity class for a given comb mode.  For all of the peaks in this group the total mode number $n_T = n_1 +n_2$ is the same; the differences arising solely from the pair of modes resonant with the transition and the intermediate state.

\subsection{Analysis Method and Frequency Extraction}\label{sec analysis}

The analysis of the data followed that detailed in Ref.~\cite{stalnaker10}.  The modeled spectra were calculated for a range of final state energy levels and hyperfine $A$ coupling constants.  The calculated spectra were fit to a multipeak fit function and the resonant repetition rate frequencies were recorded for each of the peaks in the group. For each peak the fit results were used to create an interpolating function so that the resonant repetition rate frequency of each peak could be determined for a given center of gravity energy and hyperfine $A$ coefficient.  The experimental data were also fit to a multipeak fitting function and the resonant repetition rate frequency of each peak was extracted.  The fits to the experimental data were weighted by the scaled standard deviation of the samples taken at each repetition rate frequency, where the scale was selected so that the fit returned the appropriate $\chi^2$.  The center-of-gravity frequency and hyperfine $A$ coefficient were  determined by minimizing the weighted $\chi^2$ function
\begin{align}
  \chi^2\prn{\omega_\textrm{cog},\, A} = \sum_{i=1}^N\prn{\frac{\brk{f_\textrm{r}^\textrm{expt}}_i - \brk{f_\textrm{r}^\textrm{calc}}_i\prn{\omega_\textrm{cog},\, A}}{\sigma_i^\textrm{expt}}}^2 .
\end{align}
with respect to $\omega_\textrm{cog}$ and $A$.  Here, $\brk{f_\textrm{r}^\textrm{expt}}_i$ is the resonant repetition rate frequency for peak $i$ as extracted from the fit to the experimental data, $\sigma_i^\textrm{expt}$ is the uncertainty in that value, and $\brk{f_\textrm{r}^\textrm{calc}}_i\prn{\omega_\textrm{cog},\, A}$ is the function created by interpolating the fit frequencies extracted from the modeled spectra as a function of the center of gravity frequency, $\omega_\textrm{cog}$ and hyperfine constant $A$.  The sum is over all  the $N$ peaks in the grouping that were fit.  If the uncertainties in the resonant repetition rates are normally distributed and accurately described by the fit uncertainties one would expect the minimizing $\chi^2$ function to be equal to $N - 2$.  The minimizing $\chi^2$ values were found to exceed this expected value by a factor of 2-3 for some of the scans.  This indicates the data are distributed beyond the expected noise as determined from the fit.  This additional source of scatter could arise from uncertainties in the beam misalignment or from lower frequency noise due to power variations that are not well characterized by the uncertainty in the fit positions.  The effect of beam misalignment for this type of velocity selective direct frequency comb excitation was investigated in Ref.\ \cite{stalnaker10} and was found to lead to an increase in scatter but little to no bias in the final frequency extraction.  Similarly, power fluctuations on a time scale comparable to the data acquisition time can lead to an increased scatter without biasing the data.  To account for this increased scatter, we increase the uncertainties to produce an appropriate $\chi^2$.  For each data scan we then determine the uncertainty by looking at the renormalized $\chi^2$ as a function of $\omega_\textrm{cog}$ and $A$ and finding the values where the $\chi^2$ has increased by one.  These represent the 1-$\sigma$ uncertainties in the values of $\omega_\textrm{cog}$ and $A$ for each data scan.  This procedure was done for seven different scans.  The data were then combined with a weighted mean to arrive at a final result.

\subsection{Systematic Effects}

\subsubsection{ac Stark Shifts}

Direct frequency comb spectroscopy via resonant excitation suffers from the fact that while only a few optical modes contribute to the excitation it is difficult to isolate the individual modes needed for the excitation.  Indeed, for this experiment there are more than $10,000$ optical modes present in the excitation laser beams that do not contribute to the signal but do contribute to background scattered light levels and can potentially lead to ac-Stark shifts.

Despite the large number of off-resonant optical modes the ac-Stark shifts for this experiment are dominated by the off-resonant coupling of the resonant mode to other hyperfine states.  This is a result of the $\frac{1}{\Delta}$ dependence of the ac-Stark shift, where $\Delta$ is the detuning, and the fact that the repetition rate frequency is much larger than the hyperfine structure of the intermediate state.  To understand this we note again that a shift of the intermediate states has a much smaller effect in the resonant repetition rate than a shift of the ground or excited state (Eq.\ \eqref{eq intStateSuppression}).  As a result, it is the light coupling of the 4 $P_J$ states with the ground and excited states that could potentially lead to a shift of the energy levels of the ground or excited state that is of concern.  As an example we consider excitation of the $4\: S_{1/2} \prn{F=1}\rightarrow 4\: P_{1/2} \prn{F=2} \rightarrow 6\: S_{1/2} \prn{F=1}$ transition pathway.  If a pair of comb modes are resonant with this transition, then the  mode resonant with the first stage of the transition, $n_1$, is detuned by the hyperfine splitting of the $4\: P_{1/2}$ state from the  $4\: S_{1/2} \prn{F=1}\rightarrow 4\: P_{1/2} \prn{F=1}$ transition.  This detuning is 55 MHz and is much smaller than the repetition rate frequency of 922 MHz.  As a result, the shift from the $\prn{n_1+1}$ mode contributes significantly less to the ac Stark shift as it is further detuned from the transition.  Summing over the shift from the thousands of modes present in the optical beam, taking into account the increased detuning from the resonant transitions, leads to a smaller overall shift than that of the resonant beam.  In addition to this detuning suppression, the sign of the ac-Stark shift changes with detuning.  As there are off resonant modes both above and below the resonance there will be a partial cancelation of the shift from the multiple modes.  The situation is more exaggerated for the transitions excited through the $4 \: P_{3/2}$ state as the hyperfine splitting of this state is smaller than that of the $4\: P_{1/2}$ state.

Because it is the off-resonant coupling of the resonant mode with the other hyperfine states that is most significant the shifts from the ac-Stark effect will have different signs for the transitions occurring through different hyperfine states of the intermediate state. The result will be a broadening of the overlapped peak structure corresponding to the different transition pathways for a given ground state hyperfine component and excited state hyperfine component.  While the shifts of the overlapping peaks due to the ac-Stark shift will not necessarily be of the same magnitude due to different coupling strengths, the net shift of the transition frequency will be reduced.

Given the considerations described above, we estimate the ac-Stark shift by considering only the shift arising from coupling of the resonant mode with the off resonant hyperfine states of the $4\: P_J$ states.  The largest contributor to the shift will be due to coupling of the $4\: P_{3/2}$ state to the ground state because the dipole coupling of the $4\: P_{3/2}$ to the ground state is stronger than the coupling to the excited state, the hyperfine splitting of the $4\: P_{3/2}$ state is smaller than the splitting of the $4\: P_{1/2}$ state, and the optical power per mode is comparable for the two stages of the transition.  As described above, the optical power in the 770 nm beam is $\approx 400\: \mu$W spread over 20~nm, corresponding to $\approx 11,000$ optical modes.  Assuming a uniform distribution of power over the spectral region passed though the filter and the area of the optical beam gives an intensity per mode of $\approx 0.5$ mW/cm$^2$.  The ac-Stark shift of the ground state due to the coupling of the $4\: P_{3/2}$ state is estimated to be (see, e.g.\ Ref.~\cite{budker08})
\begin{align}
  \frac{\Delta \omega_{gi}}{2\pi} = \frac{\prn{d_{gi}\, \varepsilon}^2}{4 \Delta} \lesssim  30 \: \textrm{kHz} ,
\end{align}
where $d_{gi}$ is the dipole matrix element between the ground and excited state, $\varepsilon$ is the electric field of the optical mode resonant with the transition and $\Delta \approx 9$~MHz is taken to be the characteristic hyperfine splitting of the intermediate state.  For the transitions through the $4\: P_{1/2}$ state this shift will be significantly smaller due to the larger hyperfine splitting.  This shift is less than the statistical uncertainty of the measurement.

\begin{table}
\caption{Summary of the contributions to the uncertainty budget.  All frequencies are given in kilohertz. The dominant uncertainty is due to the statistical uncertainty.  The total uncertainty is found combining the uncertainties in quadrature. } \label{tab
errorBudget}
\begin{center}
\begin{tabular}{ccc}
\hline\hline
Effect & $\frac{\omega_{cog}}{2\pi}$ & $A\prn{6\:S_{1/2}}$ \\
\hline
ac Stark & $<$ 30 & $<$ 30 \\
B-Field  & $<$ 30 & $<$ 30 \\
K Vapor Pressure & $<$ 5 &  $<$ 5\\
Statistical &  120 & 100 \\\hline Total & 130 & 110 \\ \hline \hline
\end{tabular}
\end{center}
\end{table}

\subsubsection{Zeeman Shifts}

The data were taken in the presence of the earth's magnetic field.  This field will shift the energy levels of the ground and final states due to the Zeeman effect.  The shift of a state in the presence of a magnetic field $B$ is
\begin{align}
  \frac{\Delta \omega_{gi}}{2\pi} = g_F\, M_F\, \mu_0\, B
\end{align}
where $M_F$ is the projection of the angular momentum $\vec{F}$ along the magnetic field, $\mu_0 = 14$ MHz/mT is the Bohr magneton, and $g_F$ is the Land\'{e} factor.  To estimate the effect of the Zeeman shift on the determination of the transition frequencies, we again note the insensitivity of the shifts of the intermediate state to the resonant repetition rate frequency.  Consequently, it is the ground and excited state shifts that are most significant.  In this experiment both the ground and excited states are $S_{1/2}$ states and they have the same Land\'{e} factor, $g_F = \frac{1}{2}$ (see e.g.\ Ref.~\cite{budker08}).  If the excitation light is linearly polarized the predominant effect will be a slight broadening of the transition.  The maximal shift will occur if the light is propagating along the direction of the magnetic field and is circularly polarized.  In this case, the shift in the transition frequency will be
\begin{align}
  \frac{\Delta \omega_{gf}}{2\pi} = \frac{1}{2} \mu_0\, B  \approx 300 \:\textrm{kHz},
\end{align}
where we have used $B \approx 50\:\mu$T for the magnetic field.  However, we estimate the shift to be significantly less than this since the experiment was arranged so that the dominant magnetic field present was perpendicular to the propagation of the laser light and, while the light polarization was not controlled in this experiment, it is approximately linear with any circular polarization estimated to be less than 10$\%$.  As a result, we conservatively estimate the uncertainty from the Zeeman shifts to be 30 kHz.

\subsubsection{Collisional Shifts and Temperature-Dependent Effects}

The effect of the frequency shifts due to either impurities or collisional shifts were investigated by collecting data using a second vapor cell at a significantly higher temperature, $T \approx 160^\circ$C.  These data were not used for the final determination of the frequency due to the increased uncertainties resulting from absorption effects as well as increased overlap of the transitions through the two different $P_{J}$ states.  The agreement between the data and the calculation for these data were better than $400$~kHz in the optical transition frequencies.  Using the vapor pressure densities estimated from the temperatures of the two cells and extrapolating to zero vapor pressure gives a limit of collisional shifts in the lower temperature data at a level of $<5$ kHz.

\subsubsection{Final Result}

An uncertainty budget listing the uncertainties considered in this experiment is shown in Tab.\ \ref{tab errorBudget}.  Considering all of these effects we arrive at a value for the center of gravity frequency and hyperfine $A$ coefficient for the $6\: S_{1/2}$ state of
\begin{align}
  \frac{\omega_\textrm{cog}}{2\pi} = & 822\, 951\, 698.09(13)\: \textrm{MHz} \\
  A = & 21.93(11)\: \textrm{MHz} .
\end{align}
The center of gravity frequency is in agreement with the previous measurement of
$\frac{\omega_\textrm{cog}}{2\pi} =  822\, 951\, 595(90)$ MHz \cite{thompson83,sansonetti08} and represents an improvement of a factor of 700.  The hyperfine $A$ coefficient also agrees with the previous measurements of 20.4(2.3) MHz \cite{thompson83} and 21.81(18) MHz \cite{gupta73}.

\section{Conclusions}

We have applied the technique of velocity-selective two-photon excitation to atomic potassium to measure the $4\: S_{1/2} \rightarrow 6\: S_{1/2}$ transition frequency and the hyperfine splitting of the $6\: S_{1/2}$ state.  Our results are consistent with previous measurements and have an uncertainty that is 700 times smaller for the transition's center-of-gravity frequency.  In addition, we have presented a theoretical analysis of the systematic effects present using this technique.  In particular, we have argued that the insensitivity of the technique to the frequency of the intermediate state, coupled with the step-wise excitation, leads to a reduction in the types of systematic effects typically associated with precision spectroscopy.  This work builds on that of Refs.\ \cite{stalnaker10, stalnaker12} and further illustrates the precision achievable using the technique of velocity selective direct frequency comb spectroscopy.

\section{Acknowledgements}

The authors would like to acknowledge Scott Diddams for assistance with the Ti:Sapphire oscillator, Lee Sherry and William Striegl for early contributions to the frequency comb experiment.  M.E.R.\ acknowledges funding from the Oberlin College Research Fellowship and A.N.\ acknowledges support from the Science and Technology Research Opportunity for a New Generation (STRONG) program at Oberlin College.  This experiment benefitted from funding from the National Institute of Standards and Technology Precision Measurements Grant and the the National Science Foundation under Award PHY-1305591.

\end{document}